# A Survey of Algorithms for Distributed Charging Control of Electric Vehicles in Smart Grid

Nanduni I. Nimalsiri, Chathurika P. Mediwaththe, *Member, IEEE*, Elizabeth L. Ratnam, *Member, IEEE*, Marnie Shaw, David B. Smith, *Member, IEEE*, and Saman K. Halgamuge, *Fellow, IEEE*

*Abstract*—Electric vehicles (EVs) are an eco-friendly alternative to vehicles with internal combustion engines. Despite their environmental benefits, the massive electricity demand imposed by the anticipated proliferation of EVs could jeopardize the secure and economic operation of the power grid. Hence, proper strategies for charging coordination will be indispensable to the future power grid. Coordinated EV charging schemes can be implemented as centralized, decentralized, and hierarchical systems, with the last two, referred to as distributed charging control systems. This paper reviews the recent literature of distributed charging control schemes, where the computations are distributed across multiple EVs and/or aggregators. First, we categorize optimization problems for EV charging in terms of operational aspects and cost aspects. Then under each category, we provide a comprehensive discussion on algorithms for distributed EV charge scheduling, considering the perspectives of the grid operator, the aggregator, and the EV user. We also discuss how certain algorithms proposed in the literature cope with various uncertainties inherent to distributed EV charging control problems. Finally, we outline several research directions that require further attention.

## I. INTRODUCTION

INCREASED societal awareness of environmental issues associated with vehicular emissions has spurred the development of cleaner solutions for transportation. In this respect, electrified vehicles are emerging as the defining trend of transportation [1]. Recently, a dramatic increase in the adoption of EVs has additionally been attributed to decreasing battery costs and cheaper electricity prices compared to escalating fuel prices [2]. Despite the numerous benefits of EVs, large populations of grid-connected vehicles will potentially create grid congestion problems leading to costly network expansion. For example, charging a single EV will potentially double the energy consumption of an average household [3]. In cases where millions of EVs simultaneously charging across the grid, new peak load events will arise and/or existing peak load events will be compounded.

By contrast, coordinated EV charging using intelligent control strategies supported by Information and Communication Technology (otherwise known as *smart charging*) potentially offers opportunities to improve grid utilization and limit network expansion. Researchers have been developing numerous smart charging algorithms, many of which require the acquisition and processing of large amounts of information at a central node. In cases where the large computational overhead, or requirements for supporting communication infrastructure are considered impractical, alternatives to centralized approaches, such as distributed algorithms have been considered. In particular, distributed algorithms are highly scalable both from computation and communication points of view.

As opposed to centralized control schemes where all the relevant parameters are collected and a central calculation is performed by a single entity, distributed control schemes are performed by several entities that obtain certain relevant parameters via communication. Specifically, a distributed charge control scheme assigns the processing load over several agents, so that each agent only needs to solve its own small-scale problem, and as such, each agent bears the control of the charge schedules to a certain extent. Distributed charging schemes, in particular, can be realized as *decentralized* and *hierarchical* schemes, where decentralized schemes share the computational load across EVs and hierarchical schemes share the computational load across both EVs and aggregators (intermediaries between the power grid and the EV users).

In this paper, we review a specific class of EV charging schemes, namely *distributed EV charging schemes*. The key contributions of this paper are summarized as follows: distributed EV charging has not seen a focused survey, and to the best of our knowledge, this paper is the first (1) to present an explicit review of distributed charging control algorithms; (2) to elucidate the distinction between different permutations of distributed charging control architectures, based on the method of sharing computation and the structure of communication; (3) to provide a comprehensive classification of EV charging optimization problems (OPs) to better understand the existing distributed EV charging schemes studied under operational and cost aspects of grid operators, EV users, and aggregators; and (4) to assess several distributed EV charging schemes with respect to managing uncertainties related to the power grid, the electricity market, and the behaviour of EV users.

N. I. Nimalsiri and D. B. Smith are with the Research School of Electrical, Energy and Materials Engineering, The Australian National University, Canberra, ACT 2601, Australia, and also with Data61, CSIRO, Eveleigh, NSW 2015, Australia (e-mail: nanduni.nimalsiri@anu.edu.au; david.smith@data61.csiro.au).

C. P. Mediwaththe, E. L. Ratnam, and M. Shaw are with the Research School of Electrical, Energy and Materials Engineering, The Australian National University, Canberra, ACT 2601, Australia (e-mail: chathurika.mediwaththe@anu.edu.au; elizabeth.ratnam@anu.edu.au; marnie.shaw@anu.edu.au).

S. K. Halgamuge is with the Department of Mechanical Engineering, The University of Melbourne, Melbourne, VIC 3010, Australia, and also with the Research School of Electrical, Energy and Materials Engineering, The Australian National University, Canberra, ACT 2601, Australia (e-mail: saman.halgamuge@anu.edu.au).

Digital Object Identifier 10.1109/TITS.2019.2943620

There have been several surveys related to EVs and their influences [4]–[6]. A number of surveys related to EV charging schemes are presented in [7]–[10]. By contrast, we review a specific category of EV charging schemes referred to as distributed EV charging schemes. Specifically, the authors in [7] describe centralized and decentralized schemes, without mention of hierarchical schemes that have been widely explored in the recent literature. The authors in [8] propose a more general classification of charging schemes as uncontrolled, indirectly controlled, smart, and bidirectional. In [9], charging schemes are first classified as unidirectional or bidirectional and then as centralized or decentralized, and whether mobility aspects are considered or not. Complementary to [8] and [9], we review distributed charging schemes based on two significant classifications that consider: (1) the distributed control architecture model; and (2) the objective function of the OP, and thus we present a concise and comprehensive overview of distributed charging schemes. Further, we consider several other uncertain aspects of EV charging other than EV mobility. The authors in [10] present a classification based on grid, aggregator, and customer oriented charging. By contrast, we first classify charging schemes based on the operational and cost aspects of the OP, and then further classify with regard to the objectives of grid-operator, aggregator, and EV user.

This paper is organized as follows. In Section II, we provide background to EV charging. In Section III, we identify several properties of EV charge control schemes. In Section IV, control architectures for coordinated EV charging are introduced. Section V reviews a number of distributed EV charging control schemes and related uncertainties. In Section VI, we highlight several future research directions, followed by the conclusion in Section VII.

## II. Background

An EV uses electrical energy as the principle means of propulsion. Fig. 1 illustrates the bidirectional power flow between EVs and the smart grid, where electrical energy generated by power plants and renewable energy sources (RESs) recharges the EV battery (grid-to-vehicle: G2V) and the electrical energy delivered back to the grid discharges the EV battery (vehicle-to-grid: V2G). An aggregator often acts as a proxy between EVs and the power grid as well as the electricity market in managing smart interactions, so that the grid-operator need not directly deal with a large number of EVs. In reality, an aggregator could be a utility company, an EV charging facility, a fleet operator of a parking lot or a communication device at a transformer [11]. The information exchange between smart entities is through two way digital communication enabled by ubiquitous wireless networks and broadband power lines. The power networks and communication networks together build up a complex network. Therefore, proper control strategies supported by well established communication, measurement, and control infrastructures are crucial for the successful rollout of EVs.

In an EV battery, the state of charge (SoC) is the percentage of remaining energy capacity. The relationship between the external charging power and the rate of change of SoC of the battery can be approximated using a battery model [12].

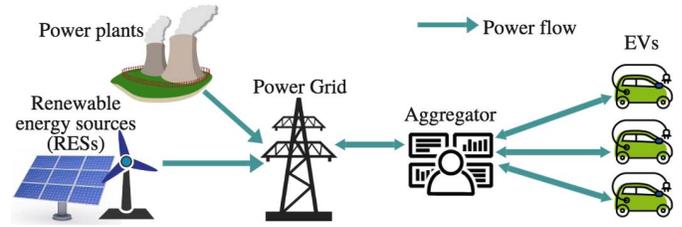

Fig. 1. Interactions between EVs and the smart grid.

Due to the conversion losses during charging, only a part of the total amount of energy drawn from the grid is effectively dispatched for charging an EV [13].

Demand-side management (DSM) of EVs refers to reducing the peak load by shifting EV charging to time periods with less congestion [14]. Given the duration that EVs are typically idle (e.g., overnight at a residence and during work hours at an office), EVs are considered an ideal prospect for DSM. As such, the three key dimensions that need to be considered for DSM of EVs are: (1) space (where to charge); (2) time (when to charge); and (3) speed (at what rate to charge). An EV owner might charge the battery overnight, e.g., start when arriving home and finish the next morning when departing for work. During the day time, an EV owner might charge the battery at the kerbside or in a parking lot. EVs can also be charged in EV charging stations (EVCSs) at times when immediate charging becomes inevitable during a trip. Compared to charging at home, EVCS operators may offer lower prices because they generally purchase large volumes of power from the wholesale power market at cheap rates. In addition, the deployment of fast charging stations, especially in densely populated areas where the majority of users have no access to over-night charging is also becoming increasingly popular. Importantly, fast charging solutions reduce long charging times and potential range anxiety of drivers. Furthermore, wireless power transfer for EVs using magnetic resonance is also gaining widespread interest [15].

EVs outfitted with bidirectional power converters (chargers and inverters) can act as electrical loads (during charging) as well as electrical sources (during discharging). Because EVs remain mostly stationary over the course of a day, opportunities to partake in ancillary services through V2G operations are possible. Nonetheless, V2G operation exhibits several drawbacks, including premature degradation of batteries and increased operational energy losses.

## III. Properties of EV Charging Control Schemes

Here we introduce some important properties relating to EV charging control schemes.

### A. One-Time, Open-Loop Versus Recursive Closed-Loop Control

One-time, open-loop control strategies (*offline strategies*) are calculated once, based on the predicted operation of the system, and thus assume perfect knowledge in advance of EV scheduling. For instance, EV charging schemes such as [13], [16], [17] are formulated such that all the EVs are available for negotiating their charge schedules at the beginning of the time horizon. However, in a more realistic setting, neither the EV arrival times nor the status of the distribution power network is known a priori. By contrast, recursive closed-loop control strategies (*online strategies*) are calculated multiple times, based on feedback measurements, hence capable of handling numerous uncertainties, including the mobility of EVs. For instance, EV charging schemes such as [18]–[20] solve the control problem progressively in the order that information becomes available (See Section V-C for more details).

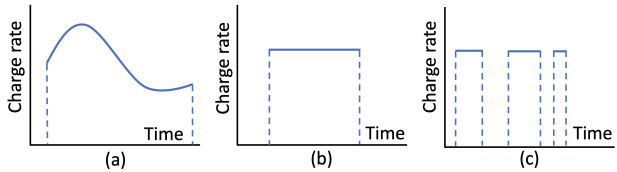

Fig. 2.   Charging at (a) variable rates and (b-c) discrete rates.

### B. Charging at Variable Versus Discrete Rates

In the literature, EV charge rate is often considered a variable that can take an infinite number of values between zero and the maximum charge rate, as depicted in Fig. 2(a) [21]–[24]. In practice, residential charging is mostly done at discrete rates (Fig. 2(b-c)), because discrete rate chargers with simple on-off controllers are much cheaper than variable rate chargers that require sophisticated equipment to modulate the power. Further, the efficiency of a charger potentially decreases if charging is not conducted at its full capacity [25]. Nevertheless, *variable rate charging (VRC)* can be better exploited for DSM, compared to charging at a fixed rate.

For an OP with VRC, it is required to find the charge rates for each time instant at which the EVs are grid-connected. In the cases of *discrete rate charging (DRC)*, the maximum output power of the charging equipment or the maximum charging power of the battery restricts the charge rate. In particular, DRC can be implemented in an uninterrupted (constant) [26] or interrupted (binary) [27] manner as shown in Figs. 2(b-c). The decision variables for the former DRC are the times at which each EV starts charging. The additional decision variables for the later DRC, where EVs are charged at discrete time slots that are separated by idle slots, are to charge at a predefined rate or to not charge at all. With such a charging approach, the battery gets to cool down during the idle slots, which in turn improves the battery life [28]. For effective operation, the number of on-off switchings should not be too high, as frequent switching will also deteriorate some batteries [29], and in cases of wide-spread implementation, frequent switching reduces the quality of power delivered to grid-connected customers [30], [31]. To balance the flexibility of VRC against the practicality of DRC, researchers have recently considered charging schemes with a finite set of charge rates between zero and a maximum value [32].

### C. Homogeneous Versus Heterogeneous EV Specifications

Certain EV charge scheduling algorithms such as the one presented in [16] perform well for a homogeneous or identical EV population. However, a practical algorithm should perform with heterogeneous charging specifications from the EV population. In particular, heterogeneity in the charge duration, charge rate, along with heterogeneity in user preferences such as charging location are needed for practical demonstrations.

### D. Pricing Strategies

A *flat* rate refers to a pricing scheme with a fixed fee for each energy unit regardless of the time at which energy is consumed. In contrast, a time-varying pricing regime provides an incentive to coordinate EV charging. For example, a *time-of-use* (TOU) rate, where electricity is billed at a different rate during peak, shoulder and off-peak periods, provides an incentive to shift EV charging from the peak pricing period. In this way, grid congestion co-incident with the peak pricing period is alleviated. However, new charging peaks (or rebound peaks) potentially form when many EVs charge simultaneously during the off-peak pricing period. To mitigate the rebound peaks, a resurgence of interest in *real-time pricing* (RTP) strategies that reflect contemporaneous power system conditions has occurred. RTP represents either the actual energy cost for a utility generating electricity or purchasing electricity at a wholesale level, or the cost imposed by a utility for load control. RTP rates are generally increasing functions of the instantaneous demand, hence users can influence the real-time electricity rate by adjusting their energy consumption [33]. In addition, *customized* pricing strategies are also proposed in certain EV charging control schemes [34], [35].

## IV. EV CHARGING CONTROL ARCHITECTURES

In this section, we describe EV charging control architectures illustrated in Fig. 3. We consider charging decisions for a group of EVs that are made by a central entity or by individual EVs, with the former called *centralized* control, and the later called *decentralized* control. *Hierarchical* control features aggregators and EVs arranged like a tree structure. An aggregator may either directly or indirectly control a group of EVs. A *direct aggregator* decides the charge schedule for each EV in the group. By contrast, an *indirect aggregator* broadcasts information signals to the EVs to coordinate their charge profiles. As such, an indirect aggregator is not required to be computationally powerful since the computation load is shared by the other entities.

### A. Centralized Control Architecture

Fig. 3(a) depicts a centralized architecture, where the charge schedule of each EV is decided by a direct aggregator, who collects the charge requirements of all the EVs, then solves an OP to determine the rate at which each EV will charge, and communicates the optimization-based charge schedule back to the EV owners. Consequently, each EV owner relinquishes some autonomy over their charge schedule. Nevertheless, centralized schemes have the advantage that they often produce optimal solutions as complete information of the entire system is available to the aggregator. Further, centralized schemes can readily consider various global system states and coupling constraints. However, such benefits must be weighed against the EV owner concerns with the privacy of information relayed to the communication network. Moreover, a single point of failure at the aggregator (e.g., failure to solve the OP) could potentially collapse the entire system, creating the need for a backup system.

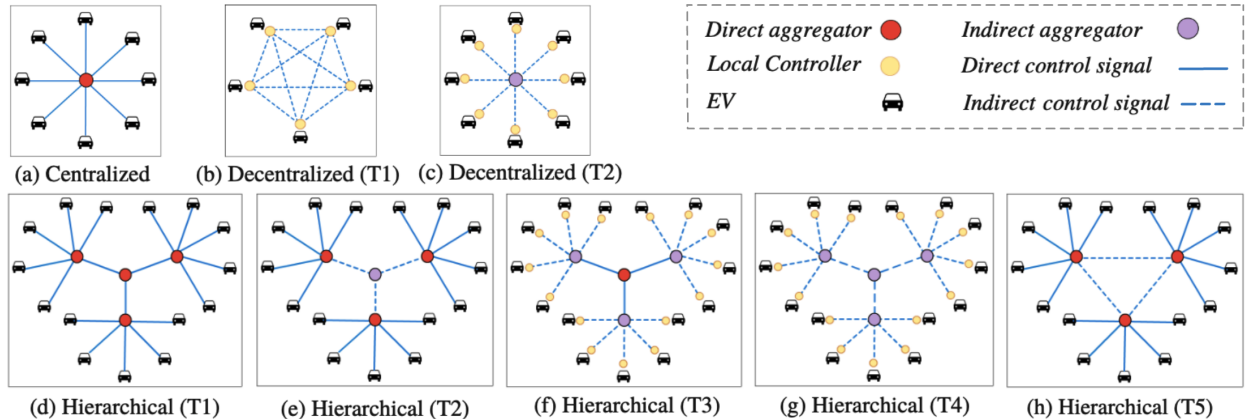

Fig. 3.   Centralized EV charging control architecture and variations of distributed (decentralized, hierarchical) EV charging control architectures.

A key challenge for centralized approaches is scalability, especially when the size of the OP increases in the length of the planning time horizon and the number of connected EVs. Therefore, a centralized approach is potentially computationally intractable with respect to the implementation time. Additional complexities arise when the number of control variables and constraints of each EV increases. Furthermore, centralized approaches require EV users to communicate to the central controller complete information of charging requirements and technical specifications of EVs. This may potentially lead to practical obstacles such as communication bottlenecks, bandwidth limitations, and costly expansion of the supporting infrastructure to handle the explosive increase of data from rapid EV uptake. Consequently, centralized approaches may lose their efficiency and become impractical when large numbers of EVs grid-connect.

### B. Decentralized Control Architecture

Decentralized systems are different from the centralized systems in that each EV acts as an independent decision-maker who solves its own problem that is small in size. As such, decentralized solutions do not always correlate with optimal charging regimes, especially in cases where there is a lack of complete information at the individual EV tier. Nevertheless, there is considerable interest in decentralized solutions, since they are highly scalable (in terms of computational complexity) and practical with respect to field implementation.

Depending on the structure of the communication network, Figs. 3(b-c) depict two decentralized control architectures. The *decentralized Type 1 (T1)* is a center free design where EVs locally compute and adjust their schedules by communicating with the other EVs, until a global equilibrium is achieved. Such an architecture enforces EVs to continuously communicate their scheduling information with the other EVs, resulting in a large communication overhead, especially when the number of EVs is very high. The *decentralized (T2)* architecture reduces the communication overhead by introducing an indirect aggregator who gathers certain information and broadcasts control (coordination) signals to all the EVs. As such, the requirement for large scale communication infrastructure is reduced. Importantly, decentralized charging schemes are more resilient to network failures, especially when controllers are designed to operate in the event of a centralized communication failure.

### C. Hierarchical Control Architecture

In the recent literature, we observe a particular interest in hierarchical control system design that is not fully centralized, nor fully decentralized. Unlike centralized systems, hierarchical systems delegate control and computational load to multiple direct or indirect aggregators via a tree-like communication topology. By doing so, the need for network-wide communication is also reduced. Each aggregator coordinates a group of EVs while influencing the decisions of the other aggregators. A group may include EVs in a single location, e.g., situated in a parking lot or located in an apartment block. Each hierarchical architecture, depicted in Figs. 3(d-h), balances the benefits of centralized and decentralized architectures in a distinctive manner. First four structures (depicted in Figs. 3(d-g)) feature three tiers: a central aggregator on the top tier, sub-aggregators in the middle tier, and EVs at the lower tier. A *hierarchical (T1)* architecture features a central aggregator that calculates a collective charging plan for all the sub-aggregators, where sub-aggregators decide each EV-specific schedule. In the *hierarchical (T2)* architecture, the central aggregator issues control signals to each sub-aggregator, transferring the computational overhead to multiple sub-aggregators that determine the charge schedules of EVs in their groups. In the *hierarchical (T3)* architecture, a central aggregator calculates a collective charging plan for all sub-aggregators, whereby each sub-aggregator indirectly controls a group of EVs by broadcasting signals, transferring the computational overhead of calculating charge schedules to the EVs. The *hierarchical (T4)* is a communication structure where all the aggregators (central aggregator and sub-aggregators) and EVs coordinate via indirect control signals. It is worth mentioning that the hierarchical (T3) and (T4) control architectures preserve decentralized behavior of EVs.

The hierarchical structures depicted in Figs. 3(d-g) are still vulnerable to single points of failure. For example, if the central aggregator collapses, all the sub-aggregators and EVs will be left uncontrolled. To mitigate against such an occurrence, the *hierarchical (T5)* architecture that is composed of a communication network across the aggregators (as depicted in Fig. 3(h)) is proposed. In cases where a link between two aggregators collapses, an alternative communication path connecting the aggregators would improve system resilience. However, if one of the aggregators collapses, EVs connected to that particular aggregator will remain uncontrolled.

## V. Distributed EV Charging Control Algorithms

A distributed algorithmic approach divides a centralized OP into a set of subproblems of a much smaller scale, that is solved by several EVs and/or aggregators. As such, decentralized and hierarchical control architectures naturally align with distributed computational systems. In this section, we review numerous decentralized and hierarchical charging control schemes that have been proposed in the literature.

Much of the early literature formulates the EV charging control problem as a *constrained OP*, with charge rates and charge durations defined as the decision variables, and with various constraints introduced to incorporate requirements of the grid operator, aggregators, and EV users. The realization of a distributed algorithm for a centralized OP is notably challenging, especially in certain non-convex OPs where the objective function and constraints are coupled to the individual charge rates of EVs, to network capacity or to electricity prices. Despite non convexity and NP-hardness, certain OPs have been solved to attain optimality or near optimality using a variety of techniques, e.g., relaxation [36], cuts [37], preprocessing [38], heuristics [19], randomization [39].

In Fig. 4, and in what follows, we organize EV charging control OPs into two categories based on (A) *Operation aspects* and (B) *Cost aspects*, and discuss numerous distributed algorithmic approaches that have been proposed to solve them.

### A. Operation Aspects

#### 1) Grid Operator's Perspective:

*a) Load regulation:* Among all possible ways of regulating the EV load, numerous studies are focused on flattening the aggregate load (EV and non-EV load) curve. By flattening load curve peaks, the risk of overloading transformers and other electrical infrastructure is reduced. Further, a flattened load curve eliminates the requirement to ramp up and down the generators, enabling a steady-state operation with maximum efficiency. Scheduling EVs to fill the overnight load valley, where the non-EV load is at its lowest, is widely addressed in the literature. Although the influence of a single EV is minor, the aggregate influence of a fleet of EVs can be substantial in terms of load flattening.

| **EV Charging Control Problems** | |
|---|---|
| **(A) Operation Aspects** | **(B) Cost Aspects** |
| Grid Operator's Perspective | Grid Operator's Perspective |
| Load regulation [16, 17, 21, 26, 41, 42, 43, 44, 51] | Minimize the cost of power operations [17, 37, 54, 85, 87] |
| Load regulation with overload constraints [3, 20, 23, 36, 46, 47, 48, 49, 50, 52, 53, 80, 91, 92, 96] | Maximize the grid operator revenue [22, 88] |
| Load regulation with voltage constraints [55, 56] | EV User's Perspective |
| Load regulation with voltage & overload constraints [54, 39] | Minimize the EV charging costs [13, 18, 22, 33, 40, 47, 58, 67, 71, 82, 86, 87, 88, 89, 90, 91, 92, 93, 94, 99, 101] |
| Maximize operational efficiency [57, 58] | |
| Aggregator's Perspective | Aggregator's Perspective |
| Manage ancillary services [34, 66, 67, 70, 71, 73] | Maximize the aggregator profit [20, 36, 40, 95, 96, 97, 98] |
| EV User's Perspective | Minimize the costs of power supply [19, 86, 99] |
| Provision for ancillary services [17, 64, 68, 69, 75, 76] | |
| Minimize the charging power losses [77, 78] | |
| Maximize the EV user convenience [24, 27, 79] | |
| Charging fairness [39, 52, 53, 57, 80, 81, 82, 83] | |
| Minimize battery degradation [18, 34, 51, 55, 71, 84, 85, 86] | |

Fig. 4. The classification of EV charging control problems and the respective distributed EV charging control schemes from the literature.

Consider a power system where $N$ number of EVs schedule their charge profiles over $T$ time slots, each of length $\Delta_t$. Let $s_i^{arr}$, $s_i^{dep}$, $b_i$, $p_i^{min}$, $p_i^{max}$ and $\eta_i$ be the SoC at arrival, the expected SoC at departure, the battery capacity, the minimum charging power, the maximum charging power, and the charging efficiency of the $i$th EV. Let $D(t)$ be the non-EV load profile, which is known a priori. The charge rates denoted by $p_i(t)$ are the decision variables. The most intuitive approach for load flattening (or valley filling) is minimizing variance of the aggregate load profile [41] and the corresponding OP is

$$\min_{p_i(t)} \quad \sum_{t=1}^{T} (D(t) + \sum_{i=1}^{N} p_i(t))^2 \tag{1a}$$

$$\text{subject to,} \quad \sum_{t=1}^{T} \eta_i\, p_i(t)\Delta_t = (s_i^{dep} - s_i^{arr})b_i, \tag{1b}$$

$$p_i^{min} \leq p_i(t) \leq p_i^{max}, \tag{1c}$$

where the energy demand and the range of acceptable charge rates of the EV are defined by constraints (1b) and (1c) respectively. One potential approach to fill the load valley is to influence EV users through electricity prices [21]. As such, an aggregator may broadcast control signals, for example, price-like signals that vary in proportion to the aggregate demand. Accordingly, each EV user will selfishly seek to minimize charging costs by scheduling the EV to charge at a time that fills the load valley.

Game theory is a promising tool that can be used to coordinate EV charging by way of optimizing individual EV charge preferences. In [16], a non-cooperative game is established to coordinate a large population of EVs who are weakly coupled via a common electricity price. The proposed game is based on a decentralized (T2) architecture, where the utility (indirect aggregator) broadcasts the aggregate EV demand after collecting the charge strategies of all the EVs. In response, EVs update their charge strategies and report them to the utility. The process iterates until the penalty imposed for the deviations of individual charge strategies from the average charge strategy vanishes. Specifically, the load curve valley is filled at the Nash Equilibrium (NE) of the game – a state where none of the EVs benefit by unilaterally deviating from their chosen charge strategies [21]. More formally,

$$f_i(p_i^*; p_{-i}^*) \geq f_i(p_i; p_{-i}^*); \quad \forall \ p_i \in P_i, \ \forall \ i \in \{1, .., N\} \quad (2)$$

where $f$ is the payoff function, $p_i$ is the charge schedule vector of EV $i$, $p_{-i}$ is the vector containing the charge schedules of all the EVs other than EV $i$ ($p_{-i} = [p_1, .., p_{i-1}, p_{i+1}, .., p_N]$), $P_i$ is the set of all feasible charge schedule vectors of EV $i$, and $p^*$ are the EV charge schedule vectors at the NE [21]. Due to nature of the specific penalization strategy in [16], the proposed algorithm is proved to be optimal only for an infinite, homogeneous EV population.

In contrast, Gan *et al.* [21] present an optimal decentralized (T2) charging (ODC) algorithm that converges to optimality for both homogeneous and heterogeneous EV fleets. In the proposed algorithm, the utility (indirect aggregator) progressively guides EVs by altering a control signal (e.g., electricity price) in response to the last received EV charge profiles. At each iteration, upon receiving the control signal from the utility, EVs individually update their charge profiles in order to minimize the sum of the electricity cost and the penalty for deviating from the charge schedule calculated in the previous iteration. Importantly, the ODC algorithm performs well even with asynchronous consumption, e.g., where EVs do not necessarily update their charge profiles in each iteration or update their charge profiles using outdated control signals. Therefore, the proposed charging scheme is quite robust to communication delays and failures. In [42], a variant of the ODC algorithm is developed to solve a discrete OP that is focused on charging EVs at discrete charge rates. In each iteration, a communication device at the transformer (indirect aggregator) broadcasts the normalized demand using the EV charge profiles from the previous iteration. Accordingly, each EV computes a probability distribution over its potential charge profiles and samples from the distribution to update its charge profile.

Based on the decentralized (T2) architecture, Li *et al.* [41] presents an online algorithm to regulate EV loads through an indirect aggregator who publishes a charging reference using the real-time aggregate EV load, after which EVs make binary decisions to charge or not, by comparing their SoC with the reference signal. It is shown that such an algorithm solves a generalized maximum weight OP. Importantly, the proposed algorithm is an on-line algorithm, which does not rely on forecasts, and as such, it is not affected by forecast errors. Using dynamic programming (DP) and game theory, an algorithm for valley filling and peak load shaving is developed in [43]. The problem of scheduling a single EV is solved using a forward induction DP algorithm. Since including multiple EVs in the DP algorithm increases the quantity of states at each time step, a non-cooperative game is formulated to coordinate multiple EVs in a decentralized (T2) manner.

A possible drawback of the mentioned charging schemes in [16], [21], [41]–[43] is the iterative nature of the respective routines, and the subsequent time they potentially take to reach the global equilibrium. Alternatively, the authors in [26] propose a non-iterative approach of scheduling a single EV at a time in a sequential manner in accordance with a decentralized (T2) model. The algorithm aims to minimize both the variance and the maximum peak of the aggregate load. A weight factor is carefully chosen to adjust the priorities between the two objectives. Once connected to the grid, an EV receives from the grid operator (indirect aggregator), the aggregate load profile for the scheduling horizon, i.e., non-EV load plus the power required to charge EVs already scheduled. Based on that information, EV solves a local OP to find its charge schedule and reports the updated load profile to the grid operator, who then submits it to the next EV. Such a scheduling approach is greedy in the sense that it determines the charge schedule of an EV only once, which occurs at the time when the EV grid-connects. Although the extensive bidirectional communication at each time step is eliminated, a possible disadvantage is the waiting period encountered by EVs that connect simultaneously. To overcome that issue, the authors modify the algorithm to update the total load profile by combining the charge schedules of all the EVs that connect simultaneously. Still, there exists the risk of forming adverse second peaks if a large number of EVs grid-connect at the exact same time. The study in [44] also utilizes a sequential scheduling approach to design a decentralized (T2) charging scheme that aims to minimize the mean square error between the real-time aggregate load and a reference operating point estimated offline using data related to non-EV load and EV mobility.

*Network-aware charging* refers to the consideration of distribution network constraints (e.g., overload control constraints, nodal voltage constraints) in EV charge scheduling. With network-aware charging, the operational envelopes of existing power systems are considered to limit costly network expansion. In what follows, we discuss load regulation schemes focused on network-aware EV charging.

*b) Load regulation with overload control:* Sustained overloading of a transformer can overheat the transformer windings, which results in its premature failure. Incorporating a distribution overload constraint as follows

$$\sum_{i=1}^{N} p_i(t) + D(t) \leq C; \quad \forall \ t \in \{1, .., T\}, \quad (3)$$

where $C$ is the rated capacity of the transformer/feeder, is a way to minimize congestion by limiting the maximum power that a transformer can carry at any time. By considering the optimal valley filling theory of ODC [21] and the distribution network topology across EVs, Ghavami *et al.* [3] develop two decentralized (T2) algorithms based on the gradient projection method (GPM), to minimize the load variance subject to the overload constraint (3). In the first algorithm, the primal problem of (1) is augmented with a cost function that associates to the overload constraint of the feeder and the GPM is applied. A limitation that exists in that algorithm is requiring the step size of the GPM to be smaller than a certain system dependent threshold. The second algorithm is an application of the primal-dual method to circumvent the nonseparability of the problem and interestingly, the second approach does not require a specific upper bound on the step size. While both overload control methods are observed to be quite effective in controlling feeder overload, the first algorithm seems to attain faster convergence whereas the second algorithm seems to perform better in terms of overload control.

By extending the algorithms in [21] and [45], the reference [23] proposes three decentralized (T2) algorithms to minimize the total load variance subject to network capacity constraint (3). Since the charge rates of EVs are coupled by both the electricity cost and the transformer capacity, two control signals are introduced to flatten the load profile and to manage the overload constraint. The first two algorithms minimize load variance using the ODC algorithm and enforce capacity constraint using the alternating direction method of multipliers (ADMM) – a powerful tool for solving large scale OPs by breaking the problem into smaller and manageable subproblems which are easier to solve. Those two algorithms differ in the time scale at which the two subroutines execute. The third algorithm employs ADMM for both subroutines. These different algorithms result in different message passing structures and exhibit different trade-offs between the feasibility and the optimality of the solutions. A sparsity promoting EV charging scheme in [46] also employs the ADMM method to generate a set of sub-problems with decoupled feeder overload constraints, which are solved by EVs independently using the dual-gradient method. However, none of the algorithms in [23], [46] are online, hence they do not capture real-time dynamics of the system.

In the decentralized (T2) charging scheme proposed in [47], the transformer overload constraints are handled by incorporating the transformer load levels in the price-signals published by the aggregator. In [48], a decentralized (T2) ant-based swarm algorithm is proposed, where EVs are treated as ants of a dynamic reunion. Each ant (EV) decides a valley filling charge schedule according to the pheromones released by the other ants, which are updated whenever the total load surpasses the maximum power capacity of the transformer. Although the convergence of the proposed algorithm is guaranteed, the time to converge is uncertain, and as such, the approach is potentially impractical for real-time implementations. Inspired by the water filling principle in information theory, the authors in [49] propose a decentralized (T2) algorithm for flattening the load profile. For each single EV, a constant water level is defined and it is adjusted through an iterative bisection method until the charge rates obtained by subtracting the non-EV load profile from the water level satisfy the energy demand of the EV. The water filling process is performed by EVs, one at a time, in coordination with the indirect-aggregator. The overload constraints of transformers are handled by reducing the energy demands of EVs according to a specific ratio, and as such, the congestion is prevented at the expense of not totally satisfying the EV demands. In order to ensure a fair dispatch of power, EVs are served in a circular order, and consequently certain EVs may encounter a considerably long waiting period. In [50], the algorithm in [49] is extended to a decentralized (T2) charging control scheme with discrete rates, using the idea of pulse width modulation. Later in [51], the algorithm in [49] is utilized to approximately solve the problem of valley filling and peak load shaving, by defining a time point before and after which EVs discharge and charge respectively.

The study in [52] presents a decentralized (T2) algorithm to avoid persistent bus congestion while ensuring a proportionally fair share of the distribution network capacity among the EVs. It is assumed that the congestion level of every branch can be measured and communicated to the downstream EVs in real-time with a reasonably low delay. To avoid bus congestion, EVs are charged at the maximum rate when the network is lightly loaded, and at a relatively low rate when the network is highly loaded. Hence, the intuition of the algorithm is to quickly control the charge rates in real-time such that the bottleneck lines and transformers are appropriately utilized. At times when the grid is overly congested, some EVs may not be fully charged by their deadlines, in favor of protecting the power system assets, thus the algorithm provides a best effort service. Since charge rates of EVs that belong to the same transformer are coupled, the dual decomposition method is used to obtain a set of distributively solvable subproblems, each of which is solved by an individual EV. Further, charge rates of EVs are adjusted based on the congestion price signals issued by the measurement nodes installed on the way to the substation. In a later work [53], the authors extend the algorithm to a dynamic setting where household loads and the number of EVs being charged change over time. The main limitation of charging schemes in [52], [53] is the heavy communication overhead, where each EV receives a message every 20ms. Moreover, the algorithms rely heavily on fast scale measurements and low latency broadband communications, hence a robust communication and measurement infrastructure is crucial.

*c) Load regulation with voltage control:* Another important consideration of a network-aware charging control scheme is the maintenance of a proper voltage level at every node of the grid. It can be enforced using an approximated power flow model [54], [55]. The authors in [55], [56] introduce an algorithm called shrunken-primal-dual subgradient, to minimize load variance while regulating nodal voltage magnitudes. The algorithm features a two-tier projection where primal and dual variables are shrunken and expanded. The system operator (indirect aggregator) guides EVs through several iterations by broadcasting the dual variable associated with the nodal voltages and the Lagrangian gradient calculated from the recent charge profiles of EVs. Unlike an ordinary primal-dual subgradient method which suffers from regularization errors, the proposed algorithm converges to optimality without regularizing the Lagrangian. However, such an algorithm requires an accurate network model and knowledge of injections and extractions of real and reactive power at every point in the network, which in practice will always be imprecise, and therefore real-time control methods to make up for these errors will be necessary.

*d) Load regulation with voltage and overload control:* By utilizing a linearized power flow model, the authors in [54] formulate an OP to regulate the charging load of EVs while minimizing the operational cost. Specifically, a network-aware charging scheme that respects both voltage and feeder transformer limits is developed. Since the transformer capacity and nodal voltages couple power flow across buses, a decoupled and decentralized (T2) algorithm is formulated using the ADMM and Frank-Wolfe methods. Besides provable convergence, the algorithm protects the privacy of users by only sharing the sum of EV charge profiles with the control center (indirect aggregator) through a communication protocol arranged over a tree graph rooted at the control center. As such, EVs which constitute the tree nodes add their charge profiles to the aggregate charge profiles from the downstream nodes and forward to the parent nodes. Upon receiving the sum of charge schedules, the control center broadcasts the cost gradient vector, based on which the EVs reschedule.

The authors in [39] present a random access charging algorithm to protect the distribution grid from bus congestion and voltage drop. The OP is formulated to maximize the number of EVs that can be charged under the given system capacity. The underlying system architecture is a decentralized (T2) model, where EVs take charging decisions based on the information of load capacities and voltage drops of system buses published by the control center (indirect aggregator). At every time slot, a set of EVs suspend charging process to provide opportunities to the waiting EVs, thereby ensures fairness within the implementation. It is worth mentioning that the respective algorithm involves no forecasts, and no convex optimizations – and as such, implementation appears to be simpler than other considered approaches.

*e) Maximize operational efficiency:* Another important operational aspect from the grid operator's standpoint is balancing the electricity generation and the demand for enhanced operational efficiency. To this end, the authors in [57] present a decentralized (T2) and token based IT infrastructure, which provides energy as a service via generation and consumption of tokens of energy. It is comprised of a heuristic algorithm to maximize the average utilization of generation, while ensuring that the total amount of power consumed by EVs is less than the amount of power allocated for charging the EVs. The practical implementation of such a system requires a perfect and seamless communication infrastructure for the negotiation of token exchange between the producers and the consumers of energy. In [58], a game theoretic approach following a decentralized (T2) architecture is proposed to balance the planned electricity generation in real-time. It is a two-stage algorithm where EVs and energy storage systems aim to flatten the load profile by adjusting the residential load to follow a day ahead energy plan. In the first stage, EV users play a non-cooperative game with mixed strategies to determine the day ahead anticipated demands that minimize the electricity costs, based on which the aggregator determines a plan to generate or purchase electricity for the next day.

In the second stage, EV owners play a real-time game to adjust their consumption patterns so as to stay close to the predicted demands. The implementation of such a scheme requires specific equipment to be installed at the houses, and hence incurs a high capital investment for the EV customers.

*2) Aggregator's Perspective:*

*Manage ancillary services:* If proper incentives are offered, EV users are likely to actively participate in a multitude of ancillary services: frequency regulation [59]–[62], voltage control [63], spinning reserve [64], [65], active and reactive power compensation [66], [67] etc. To manage the variability of renewable power generation, contribution from a large number of EVs is required. In a market environment, EVs seeking to contribute to ancillary services could be managed by a third party aggregator.

Frequency regulation is a short time scale ancillary service which aims to establish an instantaneous balance between the generation and the demand. As such, energy stored in EV batteries can be used to fine-tune the frequency and voltage of the grid by charging them when generation exceeds demand and discharging them when demand exceeds generation. A number of charging schemes are proposed to suppress the primary frequency fluctuations of the power grid [68]–[70].

Liu *et al.* [70] present a decentralized (T2), V2G control scheme, where the aggregator estimates the regulation capability of the EV fleet. Whenever the frequency deviation is out of the predefined dead band, EVs with sufficient SoC discharge power using an adaptive frequency droop control method. A potential drawback of such a method is that it requires a priori analysis of the specification of droop parameters. On the contrary, the authors in [71] propose a decentralized (T2) regulation algorithm that follows a plug-and-play concept, without requiring such parameterization. It has an indirect aggregator generating a set of virtual price signals to reflect the deviation of the aggregate EV charge/discharge profiles from the day ahead energy schedule derived from [72]. Based on these signals, EVs compute their schedules to optimize the virtual cost/income from charging and discharging.

The authors in [73] propose a backup battery bank deployed by an aggregator to maintain a stable regulation capacity. The interactions between aggregator and EVs in a V2G market are modelled as a decentralized (T2) game, where the payoff of an EV is interpreted as the payment that an EV receives for participating in the frequency regulation service. Based on the command signal issued by the grid, specifying a power level for regulation, the aggregator computes all possible Nash equilibria, selects one randomly, and then EVs simply follow it. However, the authors have not incorporated the EVs' own charging requirements into the game model.

The authors in [34] develop a V2G scheme for providing distributed spinning reserve to customers with different reliability levels. When a shortage of generation capacity or a power outage happens, customers with lower subscription of reliability are cut off, and distributed spinning reserve from EVs is used to provide power to those customers with higher subscription of reliability. The proposed decentralized (T2) scheme consists of two levels of games that are coordinated by the electricity retail market (indirect aggregator). At the lower level is a non-cooperative game that coordinates the charge schedules based on a specific V2G strategy. At the upper level, an evolutionary game is implemented to evolve the EVs' V2G strategies. For each evolving step of the evolutionary game, the lower level non-cooperative game finds a new NE. Hence, both levels reach equilibrium when the evolutionary equilibrium is reached. Although a reliability-differentiated pricing scheme appears very beneficial for realizing DSM, in practice, it is hard to differentiate the delivery of electricity in terms of reliability due to the intrinsic limitations of current power systems [74]. As a result, the research direction outlined in [34] has not received sufficient attention in the recent literature.

Active and reactive power compensation contributes to voltage regulation, power loss reduction and power factor correction [67]. The authors in [66] present a hierarchical (T1) V2G scheme for the active power compensation of EVCSs that intend to provide active power with minimum incremental costs of EVs. The aggregators (EVCS controllers) coordinate using a task-swap mechanism, such that, for each active current, a single aggregator receives the current measurement information of sudden active load change, according to which a command signal specifying the appropriate active power to be compensated is distributed to their EVs. In contrast, the authors in [67] present an algorithm for reactive power compensation of EVs. The objective function of the aggregator is interpreted as the total insufficiency of reactive power reservoir, which is to be minimized. The objective function of an EV is defined as the sum of parking cost, charging cost, and penalty cost (for unscheduled EVs), which is also to be minimized. The resultant multi objective OP is solved using the normalized normal constraint method to obtain a set of well-distributed pareto optimal solutions. Then a decentralized (T2) algorithm based on Lagrangian decomposition is proposed to make the optimization scalable as the number of EVs grows.

### 3) EV User's Perspective:

*a) Provision of ancillary services:* There is considerable literature on approaches to involve EVs in ancillary services without an intermediary, e.g., a third party aggregator. For example, the authors in [75] present a decentralized (T1) algorithm where consensus filtering is utilized to acquire consistent and accurate frequency signals by all the EVs. However, consensus mechanisms often require a large number of iterations before reaching the stopping criterion, and hence take a significantly longer computing time. In contrast, the decentralized (T2) algorithms proposed in [64], [68], [69] facilitate EVs to contribute to frequency regulation and spinning reserve according to the frequency deviation at the plug-in terminal, which is a signal of supply and demand imbalance in the power grid. The decentralized (T2) scheme proposed in [17] is comprised of two algorithms for load shifting and frequency regulation, with the later based on a frequency droop control method. Since the two control algorithms are functionally separated by time scale (load shifting on a long time scale and frequency regulation on a short time scale), they are combined to balance both objectives. The study in [76] proposes an iterative, decentralized (T2) algorithm for voltage control, where an indirect aggregator broadcasts the voltage on all pilot nodes based on the charge profiles of EVs in the previous iteration. Accordingly, EVs updates their charge profiles to limit their impact on the voltage plan.

*b) Minimize the charging power losses:* From the EV user's standpoint, another objective to consider is the minimization of power losses caused by the internal resistance of a battery during charging. The authors in [77] propose a decentralized (T1) scheme to minimize the power losses during charging, while satisfying the system constraints. The OP is characterized by a Lagrangian variable called incremental cost, and the charge rates are determined by exchanging the information of incremental cost and available global power capacity, using a consensus algorithm. A limitation of [77] is the required initialization procedure during each EV charge cycle. Later, the authors in [78] extended the work to an initialization-free charge control scheme where EVs start from any charge power allocation.

*c) Maximize the EV user convenience:* EV users often wish to attain a high level of user convenience in their charging operations. A form of an objective function for maximizing user convenience is

$$\max_{p_i(t)} \sum_{i=1}^{N} \sum_{t=1}^{T} w_i(t) \; p_i(t), \tag{4}$$

where $w_i$ is a weight factor. The authors in [27] devise a DRC scheme to select the optimal EV subset that maximizes the weighted sum (4), with a weight factor chosen to characterize the charging duration and the final SoC of EVs. The resultant combinatorial, non-convex OP is solved using the ADMM method. In the proposed hierarchical (T4) system, the sub-aggregators report the average EV energy demand of their respective groups to the central aggregator, who then completes the update of the dual variable and broadcasts to the EVs to decide whether they should charge or not based on a threshold discriminator. In contrast, Malhotra *et al.* [24] present a DRC scheme with a hierarchical (T5) control architecture to maximize the cumulative user convenience characterized by the remaining SoC, remaining time to charge, and charge rate, while also sharing the limited amount of power available from the grid (global power constraint). Here, sub-aggregators at the substations exchange EV user convenience values through a consensus algorithm and locally evaluate the specific threshold control signals to define the set of EVs allowed to charge within the global power constraint. Additionally, the local power constraints of substations are ensured by truncating the ordered set of EV user convenience values of each substation. Although the algorithm is shown optimal for the homogeneous case, it exhibits a very small optimality gap for the heterogeneous case.

In contrast to maximizing the weighted sum as in (4), another form of an objective function focused on enhancing the EV user convenience is maximizing the charge rates of EVs, such that the desired final SoC is achieved within the shortest time. The authors in [79] propose a local control method where EVs are able to act independently without relying on external control signals to operate. Specifically, each individual EV maximizes the charge rate while maintaining the service cable loading and the voltage of the customer point of connection within acceptable limits. An additional charge rate constraint is imposed to avoid large variations of the charge rate over consecutive time steps, for the purpose of prolonging the battery service time. However, when compared to the corresponding centralized control method, the proposed local control method is not as capable of maintaining network parameters within the specified limits, and thus requires larger safety margins.

*d) Charging fairness:* Classical heuristic charge scheduling algorithms ensure various types of fairness criteria, such as the first come first serve, earliest deadline first, shortest job first, lottery policy [26]. Except for the first come first serve, the other policies require centralized data collection of EVs to decide their ordering, hence can undermine the efficiency of the algorithm. Depending on the objective function chosen, the solution may provide different notions of fairness: equal access fairness, proportional fairness, max-min fairness, etc.

Inspired by the bandwidth sharing approach that is used in communication networks, the authors in [80] develop a packetized, decentralized (T2) charge control approach, where sharing power among EVs is considered analogous to the problem of sharing a constrained channel in communication systems. All the EVs are assigned with a similar automaton in order to ensure fair and equal access to the feeder capacity. EVs are then charged over multiple short time intervals using charge packets that are approved after checking whether their load could be accommodated into the distribution system. The algorithm does not require EVs to report their charge schedules back, and thus provides benefits over many other schemes by reducing communication overhead significantly.

Several charging schemes have been proposed to ensure proportional fairness based on certain priority criteria (e.g., current SoC, remaining time to charge). The authors in [81] formulate a decentralized (T1) charge control scheme to maximize the weighted sum of SoC of EVs for the next time step, using Karush-Kuhn-Tucker (KKT) conditions of optimality and consensus algorithms. Interestingly, by ensuring fairness in the SoC distribution, EVs attain a reasonable SoC even in the event of an early departure. Inspired by the concept of congestion pricing in Internet traffic control, the authors in [82] propose a charging scheme, where fairness is ensured based on the amount that the EVs wish to pay, defined in terms of a parameter called willingness-to-pay (WTP). The objective function of an EV user is chosen to maximize the difference of its utility (a non-decreasing logarithmic function of the EV demand and the WTP value) and the energy cost. An interesting observation made is that as EVs with large WTP values finish charging, the price turns cheaper for EVs with smaller WTP value.

Additive increase multiplicative decrease (AIMD) is an algorithm that is suitable for large scale systems where users join and leave frequently. AIMD also exhibits the notion of fairness by sharing the available resources fairly among all the system entities. Using the AIMD algorithm, the authors in [83] develop a set of decentralized (T2) charge control schemes to maximize the limited amount of power that can be obtained from the grid. During the additive phase of the algorithm, EVs increase their charge rates by additive factors until the aggregate power demand of EVs reaches the available power capacity, which is known as a capacity event. Upon detecting a capacity event, the multiplicative decrease phase activates and EVs reduce their charge rates by multiplicative factors determined in a probabilistic manner. Specifically, three scenarios of interest are considered, namely a domestic, a workplace, and an EVCS, with utility functions of equal power sharing, fair power sharing, and minimum time power-sharing respectively. Parameters of the domestic charging scenario are fine-tuned to assign the same priority to each EV (equal fairness), whereas the parameters of the workplace charging scenario are fine-tuned to allocate higher charge rates for EVs who are in need of more energy (proportional fairness). Compared to the algorithms which require extensive communication among system entities, AIMD is proven to be highly efficient as it can be implemented by EVs without any communication at all, except that of the notification of the capacity event.

*e) Minimize battery degradation:* In certain distributed charge-discharge schemes, as in [34], [51], [84], the minimum and maximum SoC are specified to protect the battery from early degradation. In several other schemes such as [18], [34], [55], [71], [85], [86], the cost of battery degradation is included in the objective function of the OP.

### B. Cost Aspects

Here we review cost aspects and associated objective functions for distributed algorithmic approaches from the perspective of the (1) grid operator, (2) EV user, and (3) aggregator.

*1) Grid Operator's Perspective:*

*a) Minimize the cost of power operations:* Cost of power generation is an important concern of the grid operator. Shao *et al.* [37] present a bidirectional power control framework to minimize fuel costs and startup-shutdown costs of generators. A hierarchical (T3) framework is developed to model the cooperative power dispatch among the system operator (SO) on the top level, aggregators in the middle level and EVs at the bottom level. Since the grid side formulation (upstream) and the EVs side formulation (downstream) can be connected via the aggregator net power, the system architecture fits Benders decomposition – a technique that is useful when the number of constraints of an OP is considerably high. The SO solves the master problem for determining the net power of each aggregator that contributes to a minimal overall generation cost, while also considering a series of constraints such as grid reserve, power balance, transmission capacity, and minimum/maximum aggregator net power. The resultant power shares of the aggregators and the charge/discharge powers determined by the EVs are then fine-tuned for several iterations using approximate benders cuts.

In addition to minimizing the generation cost, the algorithm in [17] aims to charge EVs with minimal carbon dioxide emissions, while reducing the dependency of conventional regulation plants. Here, the grid controller (indirect aggregator) publishes a cheap power trajectory, upon which EVs execute a decentralized (T2) algorithm that consists of two parts: gain and SoC deficiency. The former becomes significant when the due time is short and the power is cheap. The latter ensures that EVs with lower SoC receive higher charge rates. In contrast, the OP in [87] is formulated as a multi-objective OP aimed to minimize both the generator costs and the EV charging costs. In the proposed hierarchical (T2) system, the SO (central aggregator) optimizes the system dispatch using the most recent charge schedules of EVs and alters the price signal, so that the sub-aggregators reschedule. The negotiation process happens until neither SO nor sub-aggregators change decisions between two successive iterations. At this point, the system achieves a socially optimal equilibrium. In the hierarchical (T1) charge-discharge control scheme in [85], the upper-level aims to minimize the total cost of system operation by jointly dispatching generators and aggregators, and accordingly the lower-level computes the charge-discharge strategies for each EV following the dispatch instructions from the upper-level.

*b) Maximize the grid operator revenue:* Stackelberg game (SG) is a type of a non-cooperative game that deals with multi-level decision making processes of a group of followers in response to the decision of a leader [22]. For an energy trading game between the grid operator and EV groups (EVGs), Tushar *et al.* [22] model a hierarchical (T2) SG to decide the strategic electricity price that optimizes both the EV charging cost and the revenue of the grid operator from selling energy. Given the amount of energy requested by EVGs, the grid operator (leader) chooses an electricity price to maximize the revenue. Accordingly, EVGs (followers) choose the amount of energy that they wish to purchase in order to optimize a utility that captures a trade-off between the benefit from charging and the associated cost. Since the total amount of energy offered to the set of EVGs is constrained in the study, EVGs seek a variational equilibrium, which is a type of a generalized-NE (GNE) that is more socially stable. For a given GNE demands of EVGs, the grid chooses an energy price that maximizes the revenue of the grid. The game eventually converges to a socially optimal Stackelberg equilibrium, where EVGs achieve their equilibrium strategies for the optimal energy price determined by the grid operator. In [88], a decentralized (T2) charge/discharge control framework is proposed to optimize the revenue of a set of microgrids and the charging cost of EVs. Specifically, a dynamic pricing policy is proposed to decide the electricity price based on the real-time supply-demand curve of each microgrid. In response, the charging decisions are made by EVs using a multi-attribute decision process.

*2) EV User's Perspective:*

*Minimize the EV charging costs:* EV users are often considered as price anticipators who are willing to adjust their charge profiles according to their impact on the electricity price. A form of an objective function for minimizing the charging cost of a group of EVs is

$$\min_{p_i(t)} \sum_{t=1}^{T} \sum_{i=1}^{N} c(t) \ p_i(t), \qquad (5)$$

where $c(t)$ represents the electricity price, which with respect to RTP is a function of the instantaneous total demand. The intuition from equation (5) is that the action taken by a user affects the performance of the other users through $c(t)$.

In a decentralized charging setup, each EV user wishes to selfishly choose an action which minimizes its individual charging cost. As such, EV charging can be interpreted as a non-cooperative game among the EV users. If every EV user of the game picks their cost-minimizing strategy, there will be a stable state, known as the NE, where no user can decrease the cost unilaterally (2). Given the vector of charge schedules of all the other EVs other than EV $i$ ($p_{-i}$), and assuming that $p_{-i}$ is fixed, the *best response strategy* of EV $i$ can be defined by

$$\min_{p_i} \ f_i(p_i; \ p_{-i}), \qquad (6)$$

where $f_i$ is the payoff function and $p_i = \sum_{t=1}^{T} p_i(t)$. In the decentralized (T1) energy scheduling game described in [33], each user computes the best response strategy by solving (6) and announces that to the other users to update their best responses accordingly. The update process takes place until no new schedule is announced by any user. When users simply follow what is best for them, the total energy cost monotonically decreases in each iteration until the game converges to the NE. Besides provable convergence, such an algorithm is strategy proof, which means no user benefits from being untruthful when broadcasting the charge schedule. Similarly, in [47], a decentralized (T2), non-cooperative and dynamic game is proposed to coordinate EVs through price-signals that are broadcasted by an indirect aggregator.

For the problem of minimizing the overall energy cost of a set of EVs controlled by a set of aggregators, the authors in [13] devise a non-cooperative game that follows a hierarchical (T5) control architecture. For each potential subgame among the aggregators, the optimal charge profiles that result in the NE are calculated using the best response strategies (6). The significance of their study is incorporating two user behavioral models called expected utility theory and prospect theory to evaluate the ideal and non-ideal actions of the aggregators respectively. In contrast, the authors in [86] propose a non-cooperative game implemented as a decentralized (T2) system, in which the NE is calculated by considering the distribution of driving patterns and the relationship between the EV demand and its influence on the electricity spot prices. In [89], a strictly convex N-person game in the form of a decentralized (T2) system is developed using a probabilistic model of the EV charging patterns. In more detail, a data center notifies to the EVs, the mean and variance of the historical EV loads, based on which the EV users calculate the most cost-effective time to start charging their EVs for that particular day. The game continues for the next day with updated information from the previous day, and as such, each day is an iteration of the game seeking the NE state. Most importantly, smart chargers tend to learn a better strategy in a progressive manner. However, all the above-mentioned game theoretic approaches require a lot of computational overhead due to the iterative routine. In contrast, Cao *et al.* [90] propose a heuristic algorithm to minimize the charging costs in a regulated market operated under TOU pricing. Specifically, the authors consider a much realistic scenario, where the maximum EV charging power is different at various SoC levels. In addition to minimizing the charging costs (day-ahead), the authors in [91] consider minimizing the thermal overloading of transformers. The combined OP is transferred to a partial Lagrangian problem which is separable between fleet operators who coordinate with the Distribution System Operator (DSO) in a decentralized (T2) manner. The authors in [92] propose a hierarchical (T2) control scheme to minimize the EV charging costs while abiding by substation supply constraints. To avoid peak load at the substation transformer, a capacity based tariff scheme is proposed to surcharge load excursions exceeding a penalty load threshold.

None of the charging schemes mentioned above consider discharging of EVs. Interestingly, Nguyen and Song propose a decentralized (T2) algorithm to coordinate charge and discharge of multiple EVs in a building's garage. Here, EVs intend to charge and discharge in a way that the total payment to the building is minimized. Since each EV wants to schedule its energy profile in order to pay less, a non-cooperative game is played to select their best strategies (6) independently. In contrast, the authors in [93] propose a decentralized (T1) charge control system that constitutes a retail market layer of EVs and a set of aggregators who are requested to provide certain amounts of energy from EVs. In response to the prices announced by the aggregators, EVs play a multi-stage game to minimize the charge-discharge costs. The NE of the game is sought using a consensus-based algorithm, where EVs estimate the average charge available in their immediate neighborhoods and decide the amount of energy to trade.

### 3) Aggregator's Perspective:

*a) Maximize the aggregator profit:* In the real world, an aggregator is a profit-seeking entity with a large customer base. Oftentimes, aggregators purchase energy at wholesale prices through long term bilateral contracts or by participating in the day ahead electricity market based on forecasted electricity prices [94]. The OP for maximizing profits earned from selling the purchased electricity is

$$\max_{p_i(t)} \sum_{j=1}^{J} \sum_{i=1}^{N_j} \sum_{t=1}^{T} \left( c_{ret}(t) - c_{pur}(t) \right) p_i(t), \qquad (7)$$

where $i$, $N_j$, $c_{ret}$, and $c_{pur}$ denote EV $i$, number of EVs of aggregator $j$, electricity retail price, and purchase price respectively. For charging EVs in multiple local communities with multifamily dwellings, Qi *et al.* [36] propose a hierarchical (T2) scheme where a primary distribution transformer (central aggregator) serves multiple parking decks (sub-aggregators) that purchase electricity from the utility at TOU rates and sell it to the customers at retail prices. In addition to maximizing the revenue of the sub-aggregators, the unfulfilled charging demands are penalized to minimize the loss of customer goodwill. The transformer capacity constraints are enforced by applying Lagrangian relaxation and including them as penalty terms in the original objective function. The resultant OP is solved using the distributed subgradient method with respect to each parking deck. In detail, the sub-aggregators constantly report their projected power consumption profiles to the central aggregator, who then updates the Lagrangian multipliers and broadcasts to the charging decks back. Interestingly, only the aggregate charging demand of each parking deck is required to be publicized, hence the algorithm does not disclose individual EV charging information to the central aggregator.

Xu *et al.* [20] present a hierarchical (T1) charging control framework to maximize the profit of the aggregator by minimizing the energy purchase costs under TOU tariffs. A three-step procedure is carried out wherein the first step each aggregator reports to the DSO the aggregate power boundaries based on customer charging requirements and local transformer capacity limit. In the second step, the DSO decides the optimal power share of each aggregator. Then in the final step, each aggregator allocates charge rates to the EVs in their groups, using a heuristic method which chooses to dispatch power in the order of the desired SOC and the remaining parking duration of EVs. Most importantly, heuristic power allocation algorithms are often fast to compute, lending themselves to real-time operation solutions for large populations of EVs. In [95], a decentralized (T2) charging scheme is proposed, where EVs respond to the distribution locational marginal prices that are published by the aggregators, who aim to receive incentives for preventing congestion induced by the EV loads.

Setting an appropriate electricity price quote, especially at EVCSs has several trade-offs. A very high rate may turn away the customers and reduce the revenue of EVCS, whereas a very low rate may overwhelm the EVCS without earning a proper revenue. Consequently, this leads to a price competition among EVCSs under different ownership. The authors in [96] propose a decentralized (T2) framework that consists of a hierarchical game. At the upper level, a non-cooperative game models the competition between EVCSs who aim to maximize the profits by buying power at a lower price and selling them at a higher price. Based on those prices obtained from the non-cooperative game, multiple evolutionary games take place at the lower level to evolve EVs' strategies in choosing EVCSs.

EVCSs are also capable of producing their own electricity by installing renewable power generators (RPGs) and thereby earn revenue from selling electricity to both the grid and the EVs. The authors in [97] propose a decentralized (T2), supermodular, non-cooperative game to coordinate multiple EVCSs that carefully select electricity prices to maximize their revenues. If the amount of electricity generated by the RPGs is insufficient to satisfy the demand of customers, then the EVCS buys electricity from the grid at retail price. If the EVCS has excess electricity, it is sold to the grid at wholesale price. In the proposed game, a unique NE among EVCSs is achieved by playing the best response strategies. Given that the infrastructure cost incurred is not extremely high, adopting RPGs at EVCSs is considered very beneficial.

*b) Minimize the costs of power supply:* In certain schemes proposed in the literature, minimizing the costs of supplying power is considered as an objective from the aggregator perspective. For example, the authors in [86] develop a decentralized (T2) algorithm to minimize the operational cost of a utility (aggregator) and the charging cost plus battery degradation cost of the EVs, using the ADMM method. The problem is initially formulated as a joint OP with a trade-off parameter between the dual objective functions. It is then converted to a standard exchange OP with EVs and the aggregator considered as agents with coupled objective functions. At each time step, the aggregator solves its local OP and propagates incentive signals, upon which EVs schedule their charging jobs. In [98], a decentralized charging scheme is proposed to minimize the utility's cost related to active power. Specifically, a reduction of the EV charging cost is offered by the utility as an added incentive for the EV owners. By leveraging a time-dependent extension of the well known optimal power flow (OPF) problem, the authors first formulate a joint OPF-EV charging control problem, which is then solved using a valley filling bisection algorithm that is performed by EVs in a decentralized (T2) manner. However, the proposed algorithm employs time-invariant electricity prices, therefore it is not applicable for a time-varying electricity market. In [19], a hierarchical (T3) scheme is proposed to minimize the costs for electricity supply. It comprises a set of fleet agents (FAs), each of whom manages a number of contracted EVs through three main steps. In the first step, the local EV constraints are aggregated towards the FA, and in the second step, a collective charging plan for the EV fleet is determined using dynamic programming. In the final step, the FA propagates a control signal, based on which EVs locally determine their charge schedules using a heuristic function. It is interesting to note that the proposed scheme ensures a constant execution time in terms of vertical scalability (independent of the EV fleet size).

### C. Uncertain Aspects of EV Charging Control Optimization

Deterministic OPs assume that the data for a problem is known accurately in advance. However, for many practical problems like EV charging, certain information (e.g., power demand, power generation, plug-in and plug-out times of EVs, electricity prices) cannot be known with certainty. Although there is a rich literature on distributed EV charging control, not all of them have considered such uncertain aspects. In a real-world implementation, algorithms that do not adapt to these uncertain aspects are unlikely to be successful. In this section, we review how certain distributed charging schemes have managed uncertainties that are depicted in Fig. 5.

Many of the charge scheduling algorithms proposed in the literature do not consider the mobility aspects of EVs, instead, they treat EVs as static loads with fixed spatio-temporal parameters (e.g., [27], [73], [82], [94]). In contrast, a *mobility-aware* EV charge scheduling scheme adapts to various temporal variations, such as random arrivals; unplanned departures; and spatial variations that include charging locations, availability of charging slots at EVCSs, EVs' locations at different points in time along with their varying power requirements across different locations, etc. In more detail, an EV may plug-in at any random time of the day, and may plug-out before the designated deadline. Having these uncertainties present, it is not possible to follow the original charge schedule until the end of the scheduled time horizon. In order to deal with the random behaviors of EVs, many of the proposed distributed charge scheduling schemes repeat their static algorithm at the beginning of every time slot with updated information [19], [21]. For instance, in the hierarchical (T3) charging scheme [19], the adaptability to a dynamic environment is achieved through continuous repetition of the three steps of the algorithm at each time step. In contrast, the authors in [56] propose an event driven approach, which triggers the system to recompute charge sequences when one or more EVs plug-in or one or more of connected EVs plug-out before their designated deadlines. The authors in [99] model a stochastic game that embeds a Markov decision process, defined in terms of a state transition matrix that takes into consideration the randomness of EV arrivals, departures, and charging demand. Another type of an optimization method that is often used to tackle EV mobility issue is moving (receding) horizon optimization [18], [20]. This approach first finds charge schedules of EVs for a finite time horizon which begins at the current time and ends at some future time (e.g., latest departure time among all the EVs). The charge rates that are found for the current time step are executed in real, while the rest are discarded. When the next time step arrives, the control time horizon is shifted forward by one time step, and the optimization is repeated with updated information. Other than the aforementioned techniques, probability distributions can also be employed to tackle the uncertainties of EV charging with respect to the EV arrival and departure times.

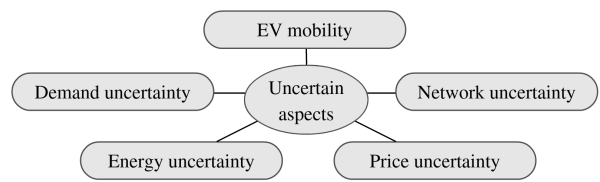

Fig. 5. Uncertain aspects of EV charging control problems.

An EV user may need to charge the battery at any location during its journey. Given the EV's route information, average speed, charging specification and location of the charging stations, the authors in [100] model a mobility-aware framework where a set of aggregators (charging service providers), each of which controls a set of charging stations, collaborate among themselves to schedule EVs that are subscribed to each aggregator in any of the charging stations of its own or others. It is interesting to note that the proposed framework is a dynamic version of the hierarchical (T5) architecture since EVs are allowed to move between aggregators.

The non-EV demand of a community often follows a trend, hence most papers consider non-EV load as deterministic and predictable ahead of time using methods like regression, time-series methods, machine learning, etc. The algorithms in [18], [92] use a similar day approach, where the loads from recent days with identical EV user behaviors and weather conditions are averaged out. Using forecasted non-EV load approximations eliminates the requirement of real-time communication and synchronization among system entities. Nevertheless, forecasted information is vulnerable to errors. Thus, it is likely that there is a discrepancy between the forecasted load and the actual power consumption. A technique that can possibly be applied to cope with demand uncertainty is recomputing the charge schedules at each new time step, with real-time non-EV load data. Alternatively, the decentralized (T2) charging scheme in [48] performs a charging adjustment mechanism to shift the charging time periods of EVs with respect to the deviation of the real-time non-EV load profile from the forecasted one.

An online charge control scheme proposed in [101] controls EV charging in each time slot entirely based on the current information, without relying on any prediction of future information. Therefore, the algorithm is robust under any EV and non-EV demand profile. It employs an event-driven discrete time model, where an event is defined by an EV arrival, an EV departure, or a change in the non-EV load profile. At any time, the scheduler computes charge sequences using the information that is available so far (charge profiles of EVs already scheduled and the past and current non-EV load) and keeps the schedules unchanged until the occurrence of the next event. As opposed to the traditional time slotted models where lengths of time slots are fixed, the time intervals of the proposed algorithm are defined by the occurrence of events, such that neither the non-EV load nor the number of EVs changes in the middle of a time interval. By doing so, the algorithm is more capable of capturing the system dynamics, which is not achievable through traditional time slotted models unless the time slots are set infinitely small.

The daily and seasonal variability of renewable energy generation contribute to energy uncertainty. The impact of energy uncertainty is particularly high if the system has a large number of small scale distributed generators, such as wind and solar. In the decentralized (T2) algorithm proposed in [17], EV charge rates are regulated to mitigate the uncertainties of electric power generation from a wind farm. The decentralized (T2) algorithm presented in [45] schedules deferrable (EV) loads to compensate for the random fluctuations in renewable generation and uncertain predictions about future power demands, which are modeled as causal filters with random deviations around their expectation. The real-time version of the particular algorithm shifts EV loads to periods with high renewable generation using information that is currently available and the updated predictions. The hierarchical (T2) SG formulated in [22] is later extended to a discrete time feedback SG in order to accommodate the time-varying nature of the amount of energy available from the grid. Further, in [83], the AIMD algorithm is exploited to assign charge rates to the EVs in almost real-time while accommodating the time varying nature of both the available power capacity and the number of EVs being charged. Another technique that is possible – although not commonly applied in the distributed EV charging context – is the Monte Carlo simulation method where the energy uncertainties can be modeled as different probability distributions.

Network uncertainty is generally related to the time-varying thermal loading sensitivities of grid components. For example, transformers and distribution feeders are sensitive to temperature and their capacities may be lower than continuous rated values on a hot day [3]. Grid components (e.g., transmission and distribution assets) may sometimes malfunction beyond a certain temperature threshold, hence their thermal capacities are required to be considered when modeling network-aware charging schemes. The thermal constraints of distribution feeders are captured in [3]. For charging control of EVs served by a single temperature-constrained substation transformer, the authors in [102] present a decentralized (T2), incentive-based and price-coordinated demand scheduling scheme that determines EV charge schedules using a dual-ascent algorithm, embedded in a predictive control scheme to introduce robustness against disturbances.

In an electricity market, price fluctuations are often a consequence of power demand fluctuations, thus price uncertainty correlates to the demand uncertainty. For example, with RTP rates, users are charged less during the valley periods and more during the peak periods. However, if users cooperate to achieve a flat load curve, then everyone will be charged by a nearly equal, and fair electricity rate, which in turn will mitigate the price uncertainty. As such, coordination of EV charging creates an opportunity for the community to save money together.

Table I summarizes several important features of distributed EV charging schemes reviewed in this paper. *Online* algorithms compute the EV charge schedules progressively, without perfect knowledge of inputs available from the start (Section III-A). The *VRC* characteristic allows for charging at variable rates (Section III-B). *Pricing scheme* indicates the type of pricing (flat rates, day-ahead, RTP, TOU or customized) used in the charging scheme, if any (Section III-D). *Mobility aware* charging considers the uncertainty of EV behaviour (e.g., arrival/departure times) (Section V-C), while *network aware* charging considers the distribution network constraints (Section V-A1). *Iterative* algorithms involve repeated computation of the charge profiles, until meeting a specific stopping criterion. Further, it is indicated whether any forecast data is used for the computation and whether V2G operations are supported by the charging scheme.

## VI. RESEARCH DIRECTIONS

In our survey of distributed charging schemes, we have presented Table I, which provides a number of useful insights. With respect to the algorithmic techniques involved, many of the distributed charging schemes require iterative calculation of the charge schedules. The computation time of such algorithms is highly affected by the time taken by an EV/aggregator to find the charge schedule/s in a single iteration, and the number of iterations is dependent on the number of participating EVs/aggregators. As such, algorithms employing less complex optimization methods to recompute the charge schedules are more practical for a large EV population. Further, many of the distributed charging schemes are driven by control signals that rely on a perfect communication medium. Preferable algorithms for real-world applications are resilient to network latency and other potential network failures, and do not require large investments for extensive bidirectional communications. Moreover, most of the reviewed distributed algorithms commonly utilize forecast data, with precise forecasting techniques deemed critical for real-world implementation.

It is evident that the problem of load regulation (load flattening) is well investigated in many distributed charging schemes related to operation aspects, however very few of them incorporate both network-awareness and mobility-awareness. In the literature for distributed EV charging, algorithms for minimizing network energy losses are not investigated satisfactorily. It is also noticeable that many papers on distributed V2G frameworks consider only one operational objective, either regulating load or regulating frequency, but not both. Hence, extensions to existing distributed V2G frameworks to achieve more than one operational objective will potentially have significant impact. Moreover, distributed EV charging schemes that consider OPs with multiple operational objectives are limited to the perspective of a single entity (e.g., they do not consider overload control and voltage control while maximizing EV user convenience). With respect to the cost aspects of EV charging control, it can be realized that distributed charging control schemes focused on enhancing the system-wide social welfare through cost optimization of multiple parties (EV users, grid operators, aggregators) are very limited.

TABLE I
CHARACTERISTICS OF EV CHARGING CONTROL SCHEMES

| Objective | Paper | Control architecture [i] | Perspective [ii] | Online | VRC | Pricing scheme [iii] | Mobility aware | Network aware | Forecasts | Iterative | Supports V2G | Technique |
|---|---|---|---|---|---|---|---|---|---|---|---|---|
| **Operation aspects** | | | | | | | | | | | | |
| Load regulation | [16] | D(T2) | G | ✗ | ✓ | - | ✗ | ✓ | ✓ | ✓ | ✗ | Game theory |
| | [21] | D(T2) | G | ✓ | ✓ | - | ✓ | ✗ | ✓ | ✓ | ✗ | GPM |
| | [41] | D(T2) | G | ✓ | ✓ | - | ✓ | ✗ | ✗ | ✓ | ✗ | Max-weight algorithm |
| | [42] | D(T2) | G | ✓ | ✗ | - | ✓ | ✗ | ✓ | ✓ | ✗ | Stochastic programming |
| | [43] | D(T2) | G | ✗ | ✓ | - | ✗ | ✗ | ✓ | ✓ | ✗ | DP+ Game theory |
| | [44] | D(T2) | G | ✓ | ✓ | - | ✓ | ✗ | ✗ | ✗ | ✗ | Arithmetic operations |
| | [45] | D(T2) | G | ✓ | ✓ | - | ✓ | ✗ | ✓ | ✓ | ✗ | GPM |
| | [51] | D(T2) | G | ✗ | ✓ | - | ✓ | ✗ | ✓ | ✓ | ✓ | Water filling |
| Load regulation with overload control constraints | [3] | D(T2) | G | ✗ | ✓ | - | ✗ | ✓ | ✓ | ✓ | ✗ | GPM |
| | [23] | D(T2) | G | ✗ | ✓ | - | ✗ | ✗ | ✓ | ✓ | ✗ | GPM + ADMM |
| | [46] | D(T2) | G | ✗ | ✓ | - | ✗ | ✓ | ✓ | ✓ | ✗ | ADMM |
| | [48] | D(T2) | G | ✓ | ✓ | - | ✓ | ✓ | ✓ | ✓ | ✓ | Ant-based swarm algorithm |
| | [49] | D(T2) | G | ✓ | ✓ | - | ✓ | ✓ | ✗ | ✗ | ✓ | Water filling |
| Load regulation with voltage and overload control constraints | [39] | D(T2) | G | ✓ | ✗ | - | ✓ | ✓ | ✗ | ✗ | ✗ | Random access algorithm |
| Load regulation with voltage constraints + Minimize battery degradation cost | [55] | D(T2) | G/E | ✗ | ✓ | - | ✗ | ✗ | ✓ | ✓ | ✗ | Primal-dual subgradient |
| Provision of ancillary services (voltage control) | [76] | D(T2) | E | ✓ | ✓ | - | ✓ | ✓ | ✗ | ✓ | ✓ | Game theory |
| Frequency regulation | [70] | D(T2) | A | ✓ | ✓ | - | ✗ | ✗ | ✗ | ✗ | ✓ | Adaptive droop control |
| | [75] | D(T1) | E | ✓ | ✓ | - | ✓ | ✗ | ✗ | ✗ | ✗ | Consensus filtering |
| | [73] | D(T2) | A | ✓ | ✓ | CP | ✗ | ✗ | ✗ | ✓ | ✓ | Game theory |
| Manage ancillary services (providing spinning reserve) | [34] | D(T2) | A | ✗ | ✓ | CP | ✗ | ✗ | ✓ | ✓ | ✓ | Game theory |
| Manage ancillary services (active power compensation) | [66] | H(T1) | A | ✓ | ✓ | - | ✗ | ✗ | ✗ | ✗ | ✓ | Task swap mechanism |
| Maximize user convenience | [27] | H(T4) | E | ✗ | ✗ | - | ✗ | ✓ | ✓ | ✓ | ✗ | ADMM |
| | [24] | H(T5) | E | ✗ | ✓ | - | ✗ | ✓ | ✓ | ✓ | ✗ | Convex optimization |
| Minimize charging power losses | [77] | D(T1) | E | ✓ | ✓ | - | ✓ | ✓ | ✗ | ✓ | ✗ | Consensus coordination |
| | [78] | D(T1) | E | ✓ | ✓ | - | ✓ | ✓ | ✓ | ✓ | ✗ | Consensus coordination |
| Charging fairness (Proportional fairness) | [83] | D(T1) | E | ✓ | ✓ | - | ✓ | ✓ | ✗ | ✓ | ✗ | AIMD |
| Charging fairness (Equal access) + Overload control | [80] | D(T2) | E/G | ✓ | ✓ | - | ✓ | ✓ | ✗ | ✓ | ✗ | Packetized approach |
| Charging fairness + Minimize charging cost | [82] | D(T1) | E | ✓ | ✓ | RTP | ✗ | ✓ | ✗ | ✓ | ✗ | Congestion pricing technique |
| Minimize bus congestion + Charging fairness | [52] | D(T2) | G/E | ✓ | ✓ | - | ✗ | ✓ | ✗ | ✓ | ✗ | Lagrangian decomposition |
| Maximize operational efficiency + Charging fairness | [57] | D(T2) | G/E | ✓ | ✓ | - | ✓ | ✓ | ✗ | ✗ | ✗ | Heuristic algorithm |
| **Cost aspects** | | | | | | | | | | | | |
| Minimize the cost of power operations | [37] | H(T3) | G | ✗ | ✓ | - | ✗ | ✓ | ✓ | ✓ | ✓ | Benders decomposition |
| | [85] | H(T1) | G | ✓ | ✓ | - | ✗ | ✗ | ✓ | ✓ | ✗ | Convex optimization |
| Minimize EV charging costs | [13] | H(T5) | E | ✗ | ✓ | RTP | ✗ | ✗ | ✓ | ✓ | ✗ | Game theory |
| | [33] | D(T1) | E | ✗ | ✓ | RTP | ✗ | ✗ | ✓ | ✓ | ✗ | Game theory |
| | [103] | D(T2) | E | ✗ | ✓ | - | ✗ | ✗ | ✓ | ✓ | ✗ | Game theory |
| | [104] | D(T2) | E | ✗ | ✓ | DAP | ✗ | ✗ | ✓ | ✓ | ✗ | Game theory |
| Minimize the power supply cost | [19] | H(T3) | A | ✓ | ✓ | - | ✓ | ✗ | ✗ | ✗ | ✗ | DP + Heuristic algorithm |
| Minimize the electricity purchase cost | [20] | H(T1) | A | ✓ | ✓ | TOU | ✓ | ✗ | ✗ | ✗ | ✗ | Linear Programming + heuristic |
| Maximize aggregator revenue | [36] | H(T2) | A | ✓ | ✓ | TOU | ✓ | ✗ | ✓ | ✗ | ✗ | Lagrangian relaxation |
| | [97] | D(T2) | A | ✗ | ✓ | CP | ✗ | ✗ | ✓ | ✗ | ✗ | Game theory |
| Minimize power supply cost + Minimize EV charging costs | [98] | D(T2) | A/E | ✓ | ✓ | Flat | ✗ | ✗ | ✓ | ✓ | ✗ | Nonsmooth separable programming |
| Maximize grid operator revenue + Minimize EV charging costs | [22] | H(T2) | G/E | ✓ | ✓ | CP | ✗ | ✗ | ✓ | ✓ | ✗ | Game theory |
| Maximize aggregator revenue + Minimize EV charging costs | [105] | D(T1) | A/E | ✗ | ✓ | CP | ✗ | ✗ | ✓ | ✓ | ✓ | Consensus coordination |
| **Operation aspects and Cost aspects** | | | | | | | | | | | | |
| Maximize operation efficiency + Minimize EV charging costs | [58] | D(T2) | G/E | ✓ | ✓ | RTP | ✓ | ✗ | ✗ | ✓ | ✗ | Game theory |
| Minimize EV charging costs + Overload control | [91] | H(T2) | E/G | ✓ | ✓ | - | ✗ | ✓ | ✓ | ✓ | ✗ | Lagrangian decomposition |
| | [92] | H(T2) | E/G | ✓ | ✓ | CP | ✗ | ✓ | ✓ | ✓ | ✗ | DP + Convex optimization |
| Frequency regulation + Minimize EV charging costs | [71] | D(T2) | A/E | ✗ | ✓ | - | ✓ | ✗ | ✓ | ✓ | ✓ | Convex optimization |
| Load regulation + Frequency regulation + Minimize the electricity generation cost and carbon dioxide emissions | [17] | D(T2) | G/E | ✗ | ✓ | - | ✗ | ✗ | ✗ | ✗ | ✗ | Linear programming |
| Load regulation with voltage and overload control constraints + Minimize cost of power operation | [54] | D(T2) | G | ✓ | ✓ | RTP | ✓ | ✓ | ✓ | ✓ | ✗ | ADMM |
| Minimize power supply cost + Minimize EV charging costs + Minimize battery degradation cost | [86] | D(T2) | A/E | ✗ | ✓ | CP | ✗ | ✗ | ✗ | ✓ | ✗ | ADMM |

[i] C: Centralized, D: Decentralized, H: Hierarchical [ii] G: Grid operator, E: EV user, A: Aggregator [iii] CP: Customized Pricing, DAP: Day Ahead Pricing

Distributed charging schemes that consider combinations of multiple objectives with respect to both the operational and cost aspects (e.g., maximizing user convenience and minimizing charging costs) are yet to be explored further. Another potential topic of interest is to study OPs where different individuals of the same entity have distinct objective functions, e.g., distributed coordination of several aggregators where certain aggregators aim to minimize the charging costs on behalf of their EV customers and the other aggregators aim to maximize their profit from selling energy to the EV customers. In addition, future research can focus on developing distributed EV charging algorithms to accommodate various uncertain aspects discussed in Section V-C.

Many of the existing distributed EV charge control schemes exploit simpler and linear battery models. However, those battery models are not accurate in practice, since the real internal power of the battery is a nonlinear function of the external power, due to internal power losses. Hence, distributed EV charge control schemes involving realistic and accurate battery models [12] are required. Moreover, the consideration of transient process of the batteries and variations in charging efficiencies is needed to improve future practical applications.

The traditional electric grid transports electricity over long distances and through complex electricity transportation routes. Alternatively, EVs can obtain electricity from other EVs through vehicle-to-vehicle (V2V) power exchange. For example, in parking lots or EVCSs, EVs with V2G capability can exchange or trade electricity in a localized peer-to-peer (P2P) manner [106]. Formulation of decentralized algorithmic approaches for supporting demand response through P2P transaction systems is another research direction of growing interest. In addition, blockchain-based, decentralized charge-discharge schemes (e.g., [107]) are preferable for managing secured and distributed energy transactions in a V2G and V2V market without reliance on a third party aggregator.

## VII. CONCLUSION

A distributed control paradigm, in contrast to a centralized approach, addresses numerous challenges (e.g., computational and communication challenges) in coordinating a large population of EVs. In this paper, we have presented a comprehensive survey of distributed EV charge control algorithms that are compatible with decentralized and hierarchical control architectures. First, we have classified OPs for EV charging control with respect to operational aspects and cost aspects. Under each category, we have reviewed the state-of-the-art distributed charge control schemes from the perspectives of the grid operator, the EV user, and the aggregator.

From the perspective of the grid operator, we have reviewed numerous distributed algorithms for load regulation, congestion management, improved efficiency, maximized revenue and minimized power generation and supply costs. With respect to the EV user, we have reviewed distributed algorithms for improved user convenience, provision of ancillary services, minimized charging losses, lower EV charging costs, and the relative fairness of different charging approaches. In considering the aggregator perspective, distributed charge control schemes with respect to managing ancillary services, maximizing the revenue, and minimizing the power supply costs are also reviewed. A crucial aspect of any EV charge control system is uncertainty. Thus, we have reviewed numerous algorithms that have been proposed to tackle various uncertain aspects of EV charging. Finally we have identified several research directions with respect to distributed EV charging control.

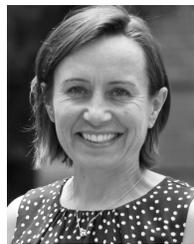
**Marnie Shaw** received the B.Sc. (Hons.) and Ph.D. degrees from the School of Physics, The University of Melbourne, Melbourne, VIC, Australia, in 1999 and 2003, respectively. She was the Head of Data Analysis at Descartes Therapeutics Inc., Boston, an Instructor at Harvard University, and a Scientist at the University of Heidelberg, Germany. She is currently a Research Leader of the Battery Storage and Grid Integration Program with the Australian National University and the Convenor of the ANU Energy Efficiency Research Cluster at the ANU Energy Change Institute. Her research interests lie in applying data analytics and machine learning to a range of data-rich problems, including the integration of renewable energy into the electricity grid.

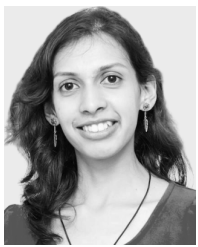
**Nanduni I. Nimalsiri** received the B.Sc.Eng. degree (Hons.) in computer science and engineering from the University of Moratuwa, Moratuwa, Sri Lanka. She is currently pursuing the Ph.D. degree in engineering and computer science, majoring in engineering, with the Research School of Engineering, The Australian National University (ANU), Canberra, ACT, Australia. She is also a Research Student with Data61, CSIRO. Her current research interests include distributed control and optimization of EV charging in smart grid.

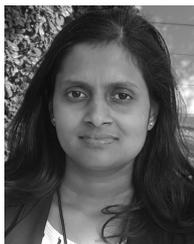
**Chathurika P. Mediwaththe** received the B.Sc. degree (Hons.) in electrical and electronic engineering from the University of Peradeniya, Sri Lanka, and the Ph.D. degree in electrical engineering from the University of New South Wales, Sydney, NSW, Australia, in 2017. She is currently a Research Fellow with the Research School of Electrical, Energy and Materials Engineering (RSEEME) and the Battery Storage and Grid Integration Program, The Australian National University, Australia. Her research interests include electricity demand-side management, renewable energy integration into distribution power networks, game theory and optimization for resource allocation in distributed networks, and machine learning.

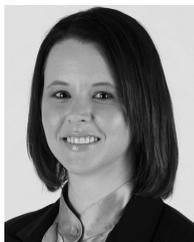
**Elizabeth L. Ratnam** received the B.Eng. degree (Hons.) and the Ph.D. degree from The University of Newcastle, Australia, all in electrical engineering, in 2006 and 2016, respectively. She subsequently held post-doctoral research positions with the Center for Energy Research, University of California at San Diego, and with the California Institute for Energy and Environment, University of California at Berkeley. From 2001 to 2012, she held various positions at Ausgrid, a utility that operates one of the largest electricity distribution networks in Australia. She currently holds a Future Engineering Research Leader (FERL) Fellowship from The Australian National University (ANU). She joined the Research School of Engineering, ANU, as a Research Fellow and a Lecturer in 2018. Her research interests include applications of optimization and control theory to power distribution networks.

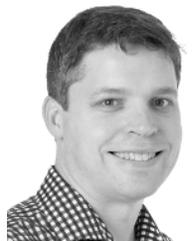
**David B. Smith** received the B.E. degree in electrical engineering from the University of New South Wales, Sydney, NSW, Australia, in 1997, and the M.E. (research) and Ph.D. degrees in telecommunications engineering from the University of Technology, Sydney, in 2001 and 2004, respectively. Since 2004, he has been with National Information and Communications Technology Australia (NICTA; incorporated into Data61 of CSIRO in 2016) and The Australian National University (ANU), Canberra, ACT, Australia. He is currently a Principal Research Scientist with Data61, CSIRO, and an Adjunct Fellow with ANU. He has a variety of industry experience in electrical and telecommunications engineering. He has published over 150 technical refereed articles. His current research interests include wireless body area networks, game theory for distributed signal processing, disaster tolerant networks, 5G networks, the IoT, distributed optimization for smart grid, electric vehicles, and privacy for networks. He has made various contributions to the IEEE standardization activity in personal area networks. He was a recipient of four conference Best Paper Awards. He is an Area Editor of *IET Smart Grid* and has served on the technical program committees for several leading international conferences in the fields of communications and networks.

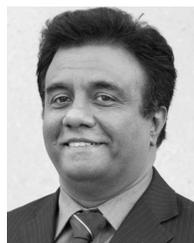
**Saman K. Halgamuge** (F'17) received the B.Sc.Eng. degree in electronic and telecommunication engineering from the University of Moratuwa, Moratuwa, Sri Lanka, and the Dr. Ing. and Dipl.Ing. degrees in electrical engineering (data engineering) from TU Darmstadt, Darmstadt, Germany, in 1990 and 1995, respectively. He has held appointments as the Director/Head of the Research School of Engineering, The Australian National University (ANU), Canberra, ACT, Australia, and the Associate Dean International of the Melbourne School of Engineering, The University of Melbourne, Melbourne, VIC, Australia, where he is currently a Professor with the Department of Mechanical Engineering. He is an honorary Professor of ANU and the ANU's Energy Change Institute. His research interests are in AI and applications in energy, mechatronics, and biomedical engineering. He has published over 250 articles and graduated 42 Ph.D. students in the above areas and obtained substantial research grants from the ARC, NHMRC, and industry. He is a Distinguished Lecturer of the IEEE Computational Intelligence Society.